Reply to Stauffer's comments.
We changed the introduction as you suggested. 
We also added the references you mention and
others of other people currently working in
econophysics. We however did not enter into 
the details of a comparison with the work of
levy-levy and solomon, mainly because it dealt
with a different system (stock exchange market) 
and a different problem (power law distributions).
Also we avoided references to the work you mentioned
of Kaufmann. We tried to be as self contained as 
possible. The question whether Nash equilibria 
have a good or bad ground state energy clearly depends
on the game one is looking at (as shown also by our 
nonlinear model where the second Nash equilibrium has 
a finite utility per player). The typical lesson 
clearly applies to situations like the tragedy of 
commons. Generally however the
ground state is not a Nash equilibrium. 

Thanks for the misprints you pointed out to us.

With respect to the parallel game theory/stat. mech.
we expanded a bit the discussion including
the issue of the symmetry of the interactions.
In the previous versions we misused the term
ground state for a state of zero temperature.
We hope now the discussion is more clear.

\documentstyle[aps]{revtex}
\newcommand{\avg}[1]{\langle{#1}\rangle}
\newcommand{\req}[1]{(\ref{#1})}
\newcommand{\beq}{\begin{equation}}
\newcommand{\eeq}{\end{equation}}
\newcommand{\beqas}{\begin{eqnarray*}}
\newcommand{\eeqas}{\end{eqnarray*}}
\begin{document}
\title{Fluctuations around Nash Equilibria in Game 
Theory}
\author{Matteo Marsili and Yi-Cheng Zhang}
\address{Institut de Physique Th\'eorique, 
Universit\'e de Fribourg, CH-1700}
\date{\today}
\maketitle

\begin{abstract}
We investigate the fluctuations induced by irrationality
in simple games with a large number of competing players. 
We show that Nash equilibria in such games are ``weakly'' 
stable: irrationality propagates and amplifies through 
players' interactions so that 
huge fluctuations can results from a small amount of 
irrationality. 
In the presence of multiple Nash equilibria, our 
statistical approach allows to establish which is the 
globally stable 
equilibrium. However characteristic times to reach this
state can be very large.
\end{abstract}

Game theory\cite{games} provides strategical thinking 
for modern economics and socio-political decisions. 
It has been an active research subject, in the
economist's community, in the past half century.
Extremely refined analysis is now being preformed,
however it was recently \cite{pw} observed that
such studies provide a far too idealised picture
of the real world in economics. One of the main 
pitfalls lies in the fact that the assumption
of rationality makes game theory deterministic.
On the other hand, we know that the economic world
is characterized by large fluctuations. These 
fluctuations are, for obvious reasons, of great interest 
for people in economics. They are becoming of
great interest also in the physicists community, where
it has been realized \cite{scaling} that economic systems 
share scaling and self organized critical behaviors
with more traditional subjects in statistical physics.

In real world irrationality is ubiquitous. This gives
us reason to use physics tools to include it in
game theory. Strikingly we find that irrationality
propagates and amplifies through players interactions
and it can lead to huge fluctuations, growing with
the number of players. This shows that irrationality
is indeed a ``relevant parameter'', which should be included
in game theory.

A game is defined as a mathematical model
for optimal strategies among competing players. 
Typically a player has a
utility function depending on 
the strategies of all the players. In this paper
we will limit ourselves to the so-called complete 
information games where
every player is aware of all other players' strategies 
and benefits. 
Under the basic assumption of rationality of all 
players, 
solutions in game theory are given by {\it Nash} 
equilibria 
of all players' strategies.
The nature of a Nash equilibrium differs qualitatively
from that of an equilibrium state in statistical mechanics.
Nash equilibria do not result from just the maximization
or minimization of some global function 
(such as e.g. 
the free energy in statistical mechanics)
but rather from the requirement that each player's strategy
must simultaneously be a local maximum with 
respect to his own strategy.
Loosely speaking, 
in game theory there is not a unique Hamiltonian, but rather
each player has his own Hamiltonian to minimize. 
The interactions among
players need not be symmetric and their
goals may be in conflict with one another.
Finally Nash equilibria gives an exact deterministic 
answer in the sense that it includes no fluctuations.
In a parallel with statistical mechanics, one could say
that game theory is a ``zero temperature'' theory.
The aim of this paper is to
include ``thermal'' fluctuations in
game theory through the Langevin approach.
For any realistic game, this issue is of utmost 
important 
since no player can have infinitely precise actions.
The ``zero temperature'' nature of game theory, 
resulting 
from complete rationality, was indeed recently 
questioned\cite{pw}.
We show that the effects of fluctuations, in standard 
games of a large number of competing players, can be 
quite 
dramatic and that they characterize the stability of
{\it Nash} equilibria.

The simplest, economy motivated model of game theory 
was introduced by Cournot in 1838 \cite{courn}:
2 firms produce quantities $x_1$ and $x_2$ respectively, 
of a homogeneous product. The market-clearing price of 
the
product depends, through the law of demand-and-offer,
on the total quantity $X=x_1+x_2$ produced: $P(X)=a-bX$.
The larger $X$ the smaller $P$ is. The model assumes
that the cost of producing a quantity $x_i$ is $cx_i$
and $c<a$. The firms choose their strategies (i.e. 
$x_i$) with the goal to maximize their 
profit (utility): $u_i=x_i[P(x_1+x_2)-c]$. 
The problem is to find $x_i$ assuming that both 
firms behave rationally.
The best response $x_1^*(x_2)$ of firm $1$ to any given
strategy $x_2$ of firm $2$ is obtained by maximizing 
$u_1(x_1,x_2)$ with respect to $x_1$ with fixed $x_2$. 
Firm $2$, assuming that $1$ behaves rationally (i.e. 
that it will play $x_1^*(x_2)$ whatever $x_2$ is) 
will choose $x_2^*$ which maximizes 
$u_2(x_1^*(x_2),x_2)$. 
This leads to $x_1^*=x_2^*=(a-c)/3b$. This solution 
highlights the essential point of the concept of 
Nash equilibrium\cite{nash}, which applies also 
to more general games. 

In a situation with $n$ players, we consider
\beq
u_i=x_iV(x_1+\ldots+x_i+\ldots+x_n).
\label{one}
\eeq
In general, one requires that $V(X)$ be a decreasing
function of $X$.
This describes, apart from a demand-and-offer law,
also situations where the gain of each
player depends on a common resource. As 
$X=x_1+\ldots+x_n$ 
grows, the resource is depleted ($V(X)$ decreases). 
$V(X)$ can eventually turn negative for $X>X_0$:
the resource has been exhausted and production gives 
rise to negative benefit for all the 
players. Generally one has $V'\sim -n^{-1}V$ 
($V'$ denotes derivative here and below) as a 
consequence 
of the fact that each $x_i$ have an effect $1/n$ on a 
global 
quantity $V$.
We shall consider $-\infty<x_i<\infty$. A negative $x_i$ 
is a quantity that, instead of being produced 
and sold, is bought by player $i$. We shall also discuss 
briefly the effects of the constraints $x_i>0$.

Technically, the Nash equilibrium is obtained by solving
\beq
\left.\frac{\partial}{\partial x_i}u_i(x_1,\ldots,x_n)
\right|_{x_j=x_i\,\forall j}=0\;\;\;\forall i\,.
\label{two}
\eeq
This equation contains the maximization 
of the utility of {\em player} $i$ and his 
expectation that all other players will do the 
same\cite{nota1}.
For the generalized Cournot model 
with $n$ firms, $a-c=1$ and $b=1/n$ (i.e. $V=1-X/n$),
eq. \req{two} gives the Nash equilibrium 
\begin{equation}
x_i=x_N\equiv\frac{n}{n+1}\;\;\hbox{and}\;\;u_i=\frac{n}
{(n+1
)^2}.
\label{nash}
\end{equation}
Note that $u_i$ is of order $1/n$: the common resource 
is 
nearly exhausted $V\simeq 0$ due to the aggressive 
strategies
$x_i\simeq 1$.

It is interesting to compare the above to the case where 
each 
player acts to maximize the total utility 
$U(x_1,...,x_n)=\sum_i u_i$.
In this case $x_i$ is given by $\partial U/\partial 
x_i=0$ and
the result is quite different: $x_i=1/2$ and $u_i=1/4$.
Strikingly the profit of each player in this case
is a factor $n$ larger than in the previous case! 

This is a typical lesson\cite{hardin} of game theory: 
when each player acts to maximize his own utility
$u_i$, the global utility is very small.
The global utility is maximized 
when all players have a common goal.
This is very similar to the dynamics in statistical 
mechanics where all degrees of freedom evolve to 
optimize 
an Hamiltonian. The maximal utility state, in spite of
being ``socially'' better (everybody behaves less 
aggressively and receives a better payoff), is 
unfortunately 
never achieved since incentives to cheat are large. 
This fact will emerge clearly from the analysis of 
fluctuations. 

In the Nash equilibrium instead, everybody is more 
aggressive (larger $x_i$) and {\it per} player benefit 
is much more meager. The crucial features which makes this
state more relevant than the social one is its
stability: The Nash equilibrium is stable because 
each player has no incentive to cheat since an 
over-aggressive move ($x_i>x_N$) would hurt the 
player himself.

It is important to note that the Nash equilibrium can be 
reached dynamically, like for example in a repeated game 
where the players adjust their strategies according to 
the gradient:
$\partial_t x_i=\partial u_i/\partial x_i$.
This observation suggests that a ``finite temperature'' 
can be included in the system, 
by considering the Langevin-like equation (in suitable
units of time): 
\beq
\partial_t x_i=\frac{\partial u_i}{\partial 
x_i}+\eta_i,
-
\label{lang}
\eeq
where $\eta_i(t)$ is gaussian
noise with $\langle\eta_i(t)\rangle=0$ and 
$\langle\eta_i(t)
\eta_j(t')\rangle=D\delta_{i,j}\delta(t-t')$.
$D$, in the statistical mechanics analogy, plays the 
role of
a finite temperature.

If $u_i$ is given by eq. \req{one}, it is possible to
find the stationary state distribution $P(\vec{x})$.
Indeed, since $V$ depends only on $X=\sum_i x_i$, it is 
convenient to perform an {\it orthonormal} 
transformation 
in the space spanned by $\vec{x}=(x_1,\ldots,x_n)$ into 
$\vec {y}=(y_1,\ldots,y_n)$ in such a way that 
$y_n=X/\sqrt{n}$. The Gram-Schmidt method\cite{linalg}
then gives $y_k=(\sum_{i\le 
k}x_i-kx_{k+1})/\sqrt{k(k+1)}$
for $k<n$.

In the new variables, the dynamics reads
\begin{eqnarray}
\partial_t y_k&=&V'y_k+\tilde\eta_k,\;\;1\le 
k<n
\label{dyky}\\
\partial_t y_n&=&\sqrt{n}V+V'y_n+\tilde\eta_n,
\label{dyny}
\end{eqnarray}
and, by orthonormality
$\langle\tilde\eta_i(t)\tilde\eta_j(t')\rangle= 
D\delta_{i,j}\delta(t-t')$.

This transformation has the virtue of displaying the 
statistical dependence of the variables in a natural 
way. 
Since $V$ and $V'$
depends on $y_n$ only, $y_n$ has a dynamics
which is independent of the $y_k$, whereas each $y_k$ is 
coupled to
$y_n$. Therefore the stationary distribution can, in 
general, be expressed as 
$P(\vec{y})=P(y_n)\prod_{k<n}P(y_k|y_n)$.
where $P(y_k|y_n)$ is the distribution of $y_k$ 
conditional to $y_n$. Eq. (\ref{dyny}) describes a 
``particle'' in a potential with thermal fluctuations
and can be solved using standard 
techniques\cite{gardiner}.
The same holds for eq. (\ref{dyky}), where $y_n$ appears
as a parameter. We find $P(\vec{y})\propto\exp[ -H/D]$ 
and
\begin{equation}
H=-\frac{V'}{2}\sum_{k=1}^{n-1}y_k^2
-\frac{Vy_n}{\sqrt{n}}-\frac{n-1}{\sqrt{n}}\int_0^{y_n}d
xV(\sqrt{n}x)
\label{H}
\end{equation}
This form of the stationary distribution is 
reminiscent of an equilibrium system with 
Hamiltonian $H$. It would however be misleading
to identify $-H$ with some measure of the utility 
$U$. This stationary state has a completely
dynamic origin.

The equilibrium distribution, for $D$ small, can be 
expanded around 
its maximum. The maximum of $H(\vec{y})$ is attained at
$y_k^*=\delta_{k,n}\sqrt{n}x_N$, in agreement with eq. 
(\ref{nash}).
The gaussian fluctuations around the Nash equilibrium
are found in the standard way: Expand $H(\vec y)$ up to 
second order in $\delta y_k=y_k-y_k^*$. The inverse of 
the matrix of the quadratic form, yields the 
fluctuations
$\avg{\delta y_k\delta y_j}$. In view of eq. \req{H},
one finds $\avg{\delta y_k\delta 
y_j}=\delta_{k,j}D/|V'|$ 
for $k<n$. Note that since $|V'|\sim n^{-1}$, these 
fluctuations are of order $n$. Because of these huge 
fluctuations, we shall call $y_k$ ``soft modes''.
The fluctuations of $y_n$ instead turn out to be of 
order
$1$. One can infer the fluctuations of the $x_i$'s by
using the identity
\beq
\sum_{i=1}^nx_i^2=\sum_{k=1}^n y_k^2
\label{ident}
\eeq
and assuming $\avg{\delta x_i\delta 
x_j}=(A-C)\delta_{i,j}
+C$. We discuss here only the case $V=1-X/n$ which 
allows
more compact expressions. The same features discussed 
below
apply to any $V(x)$ such that $V'\sim -n^{-1}$.
Observing that $\avg{y_n^2}=A+(n-1)C$, and using eq.
\req{ident} one finds 
\beq
\avg{\delta x_i^2}=\frac{2n^2+n-2}{2(n+1)}D,\;\;\;
\avg{\delta x_i\delta x_j}=-\frac{2n+1}{2(n+1)}D
\label{corrx}
\eeq

The main message of eq. (\ref{corrx}) is that 
fluctuations around 
the Nash equilibrium are very strong: 
The relative fluctuation of $x_i$, given that
the avergage
of $x_i$ is close to one, is proportional
to $\sqrt{nD}$. This depends on 
the fact that
$|V'|\sim n^{-1}$, which is a very general feature in 
large
games of the form \req{one}. The variable $x_i$  
fluctuates the same order of magnitude 
as 
the sum of $x_i$ over all 
$i=1,\ldots,n$.
This is possible because of the negative 
correlation
among the variables. A fluctuation of one of the 
variables 
is compensated by opposite fluctuations of the others.

Let us see what happens to 
the total utility. This is best seen
in the variables $y_k$, because 
$U=\sqrt{n}y_nV(\sqrt{n}y_n)$
depends only on $y_n$. Therefore expanding $U$ up
to second order around $\vec{y}^*$ and taking the 
average, we find
\beq
\langle U\rangle =\frac{n^2}{(n+1)^2}-\frac{nD}{n+1},
\label{Ufluc}
\eeq
where, again we assumed $V(X)=1-X/n$.
As can be easily seen, the fluctuations decrease 
the utility by a term $\delta U\simeq -D$
and they can have a dramatic effect:
If $D>D_c\simeq 1$ the average utility becomes 
negative! 

With respect to the dynamics, it is easy to check that 
the 
correlation function of the ``soft'' modes $y_k$, in the 
steady state for the linear $V(x)$, is 
\begin{equation}%
\langle y_k(t)y_k(t+\tau)\rangle \simeq nD\exp(-\tau/n).
\end{equation}%
which implies very long correlation times in the 
stationary state. This applies to the correlations 
of $x_i$ as well.

The features discussed thus far hold the same if
the constraint $x_i>0$ is imposed. The fluctuation 
around 
the Nash equilibrium eq.(\ref{nash}), at the level of 
the 
Gaussian approximation are still given by the above 
results. 
These characterize correctly 
the neighborhood of the Nash equilibrium.
The corrections to the gaussian fluctuations are 
negligible 
when $\delta x_i$ is much less than $x_N$, which occurs 
for 
$D\ll n^{-1}$. Numerical simulations show that, 
even for larger $D$, the same qualitative features 
(large fluctuations and eventually $U<0$) hold also 
in the presence of the constraint $x_i>0$.

It is instructive to study the ``social'' equilibrium
in the same way. Now each player attempts to maximize 
the total utility $U$, and the Langevin equation is 
$\partial_t x_i=\partial U/\partial x_i+\eta_i$.
In the variables $\vec y$ we find:
$\partial y_n=\sqrt{n}V+ny_nV'+\tilde\eta_n$ and 
$\partial y_k=\tilde\eta_k$ for $k<n$.
Note that $y_k$ {\it now behave as random walks}.
The distribution of $y_k$ at long times
is $P(\vec y,t)\propto \exp [-y_k^2/(2Dt)]$
and the correlations are $\langle y_k^2\rangle = Dt$
for $k<n$ and $\langle\delta y_n^2\rangle \sim D$.
This implies unbounded fluctuations of $x_i$
(i.e. $\langle\delta x_i^2\rangle \simeq Dt$)
and a negative correlation $\langle\delta 
x_i\delta x_j\rangle/\langle\delta 
x_i^2\rangle\to -1/(n-1)$ such that the 
fluctuations of the sum $X$ are finite.
This implies that the average utility 
remains finite. 
The absence of a stationary distribution in the 
``social equilibrium'' reflects its instability.

The results generalize with little qualitative
changes when one considers a more general correlation
among $\eta_i$ or a mixed ``social''
egoistic model. These and other generalizations, 
as well as more detailed calculations, will be presented 
in
a forthcoming publication. 

It is generally recognized in economy that
in realistic situations the relation between the utility
and wealth (net profit) is not linear\cite{pfolio}. 
One source of non-linearity, for example, is 
inefficiency 
in capital management. This can be tolerated by the rich 
whereas it is very dangerous for the poor. 
Most studies\cite{pfolio} assume empirically 
a quadratic relation \cite{toul}. This, assuming $x_iV$ 
as 
a measure of the wealth of player $i$, leads to
\beq
u_i=x_iV(X)[1-rx_iV(X)].
\label{uinl}
\eeq

Let us define $\bar x=X/n$ and assume that $V=1-\bar x$ 
(with little loss of generality since non-linearities 
in $V$ do not change qualitatively the results).
The interesting feature of this model is that 
if $r>2$ a new Nash equilibrium appears. Indeed
\beq
\left.\frac{\partial u_i}{\partial x_i}\right|_{x_i=\bar 
x}
=\left(1-\frac{n+1}{n}\bar x\right)\left[1-2r\bar 
x(1-\bar 
x)\right]=0,
\eeq
for $r\ge 2$, has three solutions: $\bar x=x_N$, the 
usual Nash equilibrium eq. \req{nash}, 
$\bar x=x_r=(r-\sqrt{r(r-2)})/2r$,
which is a new Nash equilibrium, and 
$\bar x=x_+=(r+\sqrt{r(r-2)})/2r$, which is an unstable
equilibrium (i.e. a minimum of the utility). 
Provided $2< r <n/2$, one
has $x_r<x_+<x_N$. The utility in the new equilibrium is
$u_i(x_j=x_r,\forall j)=1/(4r)$
which is positive and finite as compared to that at 
$x_N$ which is $O(1/n)$. 
In the presence of two equilibria a player
will choose one or the other according to what he judges
other players will do. 

Situations with more than one stable
solution are frequent and of great interest in economy
\cite{pw}. In particular one would like to know under 
what conditions a state is selected. 
The framework of Langevin dynamics \req{lang}
is particularly appealing. Indeed as we shall see it 
allows to understand which state is globally stable and 
how long a transition from the other state into it will 
take.

\widetext
From eq. \req{lang} we can derive the equation for $\bar 
x$:
\beq
\partial_t\bar x=\left(1-\frac{\bar x}{x_N}\right)
[1-2r\bar x(1-\bar x)]
+\frac{2r(1-\bar x)}{n^2}\left(\sum_{i=1}^nx_i^2-n{\bar 
x}^2\right)+
\bar\eta.
\label{nl2}
\eeq
Here $\bar\eta$ is the average of $\eta_i$, i.e. it is a
white noise with equal time correlation $D/n$.
In spite of the fact that all $x_i$ appear in eq. 
\req{nl2}, it is still useful to use the variables 
$y_k$. 
Indeed one can use the identity \req{ident}
and average over the degrees of freedom $y_k$ 
in eq. \req{nl2}. This amounts to replacing the term in 
brackets by $(n-1)\avg{y_k^2|\bar x}$ \cite{nota2}
which is the average of $y_k^2$ conditional to a
fixed $\bar x$.
In order to close the equations, we assume that in the 
Langevin equation for $y_k$,
\[\partial_ty_k=-\left[2r(1-\bar 
x)^2+\frac{1}{n}\right]y_k
+\frac{2r(1-\bar x)}{n\sqrt{k(k+1)}}
\left[\sum_{i=1}^kx_i^2-kx_{k+1}^2\right]
+\tilde\eta_k.\]
\narrowtext
\noindent
we can neglect the second term in the right hand side. 
This can be justified by the expectation that this term
is negligible for $n\gg 1$ if $x_i^2\approx x_j^2$.
The equation for $y_k$ then simplifies considerably 
and one finds that, in the steady state, 
\beq
\avg{y_k^2|\bar x}=\frac{Dn}{4nr(1-\bar x)^2+2}.
\label{yk2}
\eeq
This, in eq. \req{nl2}, gives (to leading order in $n$)
\beq
\partial_t\bar x=\frac{x_N\!-\!\bar x}{x_N}
[1\!-\!2r\bar x(1\!-\!\bar x)]+
\frac{rD(1\!-\!\bar x)}{4nr(1\!-\!\bar 
x)^2\!+\!2}+\bar\eta,
\label{nl3}
\eeq
which can be cast in the form $\partial_t\bar 
x=-dH/d\bar x+
\bar\eta$, where $-H(\bar x)$ is the integral of the 
deterministic part of eq. \req{nl3}. 
Since $\avg{\bar\eta(t)\bar\eta(t')}=(D/n)\delta(t-t')$,
the stationary solution is $P(\bar 
x)\propto\exp[-nH(\bar x)/D]$.
Here $H$ plays the role of a free energy. Indeed it has 
the form $H=E-TS$ where $T=D/n$ is the analogous of the 
temperature and $S$ is the entropy. The entropy enters 
from the
fluctuations of the degrees of freedom $y_k$ which have
been self consistently retained in the equation for 
$\bar 
x$. Let us discuss, from this point of view, the 
statistics
of $\bar x$ in the steady state: Fixing $E(x_+)=0$, the 
energy 
in the two equilibrium states are, to leading order in 
$n$,
\beqas
E(x_N)&=&\frac{2r\!-\!3\!-\!4r(r\!-\!2)x_r}{24r}
+O(n^{-1})\\
E(x_r)&=&\frac{(r\!-\!2)(1\!-\!2x_r)}{6}+O(n^{-1})
\eeqas
Therefore $E(x_N)<E(x_r)$ in the interval $2\le r< 9/4$
whereas $E(x_N)>E(x_r)$ for $r>9/4$.
It is important to stress that this energy cannot be 
interpreted as $-U$. Indeed note that the minimum
energy is at $x_N$ for $2\le r< 9/4$, whereas
the maximum utility $U$ is always at $x_r$.

Energy alone suggests therefore that 
the system will fall in the minimum energy minimum
for $t\to\infty$, and this, in the limit
$n\to\infty$, is $x_N$ for $ r< 9/4$ and 
$x_r$ for $ r>9/4$.
This conclusion holds to leading order in $n$
even if one considers also the entropy. The 
reason is that, for $n\to\infty$,
one is considering very small temperatures
$T=D/n$. Direct calculation shows that $S(x_N)=(\log 
n)/4
+O(n^{-2})$ while $S(x_r)=[\log\left(2rx_r-1\right)]/4
+O(n^{-1})$.
The entropy in $x_N$ is considerably larger
than that in $x_r$. Indeed the fluctuations of $y_k$ are 
of order $\sqrt{n}$ 
in $x_N$, whereas in $x_r$ they are finite [see eq. 
\req{yk2}].
In other words, since $\avg{\delta x_i^2} \simeq 
\avg{y_k^2|\bar x}$,
the set $\{x_i\}$ is much more widely spread in the 
$x_N$ 
minimum than in the $x_r$ one. And this is an effect 
which is
correctly accounted by the entropy above.

Even though we can identify a globally stable 
equilibrium
($x_N$ for $r<9/4$ and $x_r$ otherwise) which will 
ultimately
attract the system under the Langevin dynamics, it
is important to stress that the other {\em metastable} 
equilibrium can be stable over times which are 
exponentially
large in $n$. Indeed the energy barrier between the two 
minima 
is finite, but the temperature is very small $T=D/n$. 
If $r>9/4$ and initially the system is in the $x_N$ 
equilibrium, 
it will not visit the state $x_r$ before a time of the 
order 
of $\sim \exp[nE(x_N)/D]$. This time can be infinite
for all practical purposes. In other words, the system 
is very sensible to initial conditions.

We have presented a general approach to extend game 
theory to include fluctuations. We used the Langevin 
formulation which provides a natural bridge between 
game theory and statistical mechanics. The essential 
difference is that individual utility functions replace 
a global Hamiltonian. Fluctuations describe in a natural
way the stability nature of various equilibria. 
We find that the Nash equilibrium of simple games
with competition is stable against thermal
fluctuations, even though the amplitude of fluctuations
is very large. On the contrary, the ``socially ideal'' 
state is marginally unstable due to the presence of 
``soft
modes''. The approach also allows to study situation 
with more than one Nash equilibria and identifies the
globally stable one as well as the criteria under 
which a state is reached by the dynamics.

This work was supported by the Swiss National Foundation 
under grant 20-40672.94/1.


\begin{references}
\bibitem{nota1} Eq. \req{two} assumes that the model
is symmetric under interchanges $i\leftrightarrow j$. 
If this is not so, $x_j$ must be set to the best
response $x_j^*$ to all other players' strategies.
\bibitem{nota2} We assume that the statistics 
		of $y_k$ is the same $\forall k<n$.
\end{references}
\end{document}